# From random to directed motion: Understanding chemotaxis in E. Coli within a simplified model

**Melissa Reneaux§** and **Manoj Gopalakrishnan¶\***,

§ Department of Physics, St. Stephen's College, University Enclave, Delhi 110007, India.

¶ Harish-Chandra Research Institute, Chhatnag Road, Jhunsi, Allahabad 211019, India.

**Abstract**

**The bacterium E.Coli swims in a zig-zag manner, in a series of straight runs and tumbles occurring alternately, with the run-durations dependent on the local spatial gradient of chemo-attractants/repellants. This enables the organism to move towards nutrient sources and move away from toxins. The signal transduction network of E.Coli has been well-characterized, and theoretical modeling has been used, with some success, in understanding its many remarkable features, including the near-perfect adaptation to spatially uniform stimulus. We study a reduced form of this network, with 3 methylation states for the receptor instead of 5. We derive an analytical form of the response function of the tumbling rate and use it to compute the drift velocity of the bacterium in the presence of a weak spatial attractant gradient.**

**Introduction**

Procaryotic microorganisms like bacteria have to keep sensing their chemical environment to survive, and are able to adjust their motility in response to changes in it (Bray, 2001). This ability to bias the motion towards favourable stimuli (*attractants*, eg. Oxygen, nutrients) and away from unfavourable ones (*repellants*, eg. toxins) is referred to as *chemotaxis*. In particular, chemotaxis in the common bacterium *E. Coli* has been well-characterized. In a neutral solution, devoid of attractants/repellants, E. Coli swims in a zig-zag manner, in a random walk. In the presence of an attractant, however, the walk becomes biased towards the source of the attractant (the reverse happens in the case of a repellant) (Berg, 2003).

How does E. Coli bias its motion? Though the abrupt switch in direction during a straight swim might appear to be a stochastic process, it is not really so; rather, this process is regulated by the signal transduction machinery. E. Coli senses the attractants through receptor proteins which exist as a single cluster at one pole of the cell. The two main types of receptors are Tar and Tsr. The receptor protein is linked to the protein kinase CheA through the linker protein CheW, and these three are believed to function as a single signaling complex, which exists in active or inactive state, and undergoes stochastic switching between the states (Asakura and Honda, 1984). In the active state, CheA undergoes auto-phosphorylation and transfers the phosphoryl group to CheY, which diffuses through the cytoplasm and functions as a response regulator (McNab and Koshland, 1972; Koshland, 1977). CheY binds to the protein FliM at the base of the flagellar motor and increases the rate of switching from the CCW to CW mode of rotation. When one or more flagella thus reverses the sense of rotation, the bacterium tumbles over(Block et. al., 1982; 1983).

---

\* Corresponding author. **Present address:** Department of Physics, Indian Institute of Technology Madras, Chennai 600036, India**, e-mail:** manoj@physics.iitm.ac.in

The presence of attractants in the solution modifies the chain of events described before. When an attractant molecule binds to a receptor protein, its probability of being active is reduced; therefore, the phosphorylation of CheY is adversely affected, and the frequency of tumbles is reduced. The manifestation of this is that the bacterium tend to spend more time swimming straight without a change of direction. But how does it ensure that it will move towards the source of the attractant? And, how is the motion affected in the presence of a uniform concentration of attractant, as opposed to a spatial gradient? These questions can be answered only by considering another important component of the signal transduction pathway, i.e, methylation and de-methylation processes of the receptor.

The common receptor Tar has five methylation states, with a maximum of four and a minimum of zero methyl groups per receptor. The probability that a receptor-CheA complex is active increases with the methylation level (almost linearly at low attractant concentrations, see Sourjik and Berg, 2002; Kollman et. al., 2005). Methylation of the receptor is accomplished through the protein CheR, and demethylation is done by the phosphorylated form of another protein, CheB. The phosphorylation of CheB is done by active CheA itself, which provides an effective negative feedback in the chain: when attractant binding lowers the mean receptor activity, phosphorylation of CheB is reduced and therefore the mean methylation level goes up, which increases the activity. As a consequence, in the presence of a uniform attractant concentration, a steady state is reached and the system adapts. In the case of E. Coli, the adaptation to methyl aspartate, a common attractant is near-perfect over five orders of magnitude of the background concentrations.

In the presence of spatially varying stimulus, ligand binding and changes in methylation compete with each other, and methylation being a slower process, the receptor activity is consistently reduced over time. Therefore, the tumbles become less frequent whenever the direction of a straight swim (a vector) has a non-zero component along the direction of the attractant gradient. This results in a directionality of motion over several tumbles, and manifests itself as a drift in the direction of the attractant gradient, towards its source. If $L$ is the local chemo-attractant concentration and $\nabla L$ its spatial gradient, then for small gradients, we expect that the drift velocity increases proportional to the gradient:

$v_d \approx \kappa(L)\nabla L$ [1]

In this article, we present a preliminary calculation of the drift velocity of motion of a single bacterium, at small attractant concentrations, using the known features of the underlying biochemical circuitry. A more detailed and extended version of these results will be published elsewhere in future.

**Methylation-demethylation reactions**

The mathematical description of methylation-demethylation reactions presented here is mostly based on the formalism presented in Emonet and Cluzel, 2008, which itself is based on several earlier papers(Barkai and Leibler, 1997; Morton-Firth et. al., 1999; Sourjik and Berg, 2002; Melo and Tu, 2003; Kollman et. al., 2005). However, our notations differ from that in Emonet and Cluzel, 2008 in many cases.

Let us denote by $X_m$ the fraction of receptor complexes with m methyl groups, and let $r$ denote the methylation rate and $b$ denote the demethylation rate of the complex. Let $a_m$ be the fraction of active receptors with m methyl groups. We now assume, in conformity with the assumptions of the Barkai-Leibler model, that CheR binds only to inactive receptors and CheB binds only to active receptors. Then, the dynamical equations for methylation-demethylation reactions can be written in the form

$$\frac{dX_0}{dt} = -r(1-a_0)X_0 + ba_1 X_1$$
$$\frac{dX_m}{dt} = r[(1-a_{m-1})X_{m-1} - (1-a_m)X_m] + b[a_{m+1}X_{m+1} - a_m X_m] \quad \text{for } 1 \leq m < m_{\max} \quad [2]$$
$$\frac{dX_{m_{\max}}}{dt} = r(1-a_{m_{\max}-1}) - ba_{m_{\max}} X_{m_{\max}},$$

where the rates $r$ and $b$ have the Michaelis-Menten forms

$$r = \frac{\omega_r R_0}{K_R + A} \quad \text{and} \quad b = \frac{\omega_b B'}{K_B + A^*} \quad [3]$$

where $\omega_r$ and $\omega_b$ are the rates of formation of the final products (methylated and de-methylated receptors) from the inter-mediate complexes of receptor-CheA with CheR and CheB-P respectively, and $B'$ denote the total concentration of CheB-P. $R_0$ is the concentration of CheR, which is assumed to be large, so depletion effects are neglected. $A$ and $A^*$ denote the concentrations of inactive and active receptor-CheA respectively, and are normalized as $A + A^* = A_0$, with $A_0$ being the total concentration of receptor-CheA in the cell (assumed equal to the concentration of CheA).

In order to complete the reaction scheme, we also need to consider the kinetics of CheB-P. This is described by the equation

$$\frac{dB'}{dt} \approx a_P k_P A^*(t) B_0 - d_b \frac{K_b}{K_b + A^*(t)} B'(t) \quad [4]$$

where $a_p$ is the fraction of active receptors in phosphorylated state, $B_0$ is the total concentration of CheB in solution (also assumed large), $k_P$ is the second order association constant for the binding of phosphorylated CheA and CheB and $d_b$ is the dephosphorylation rate of CheB-P in solution. The extra factor $K_b/(K_b + A^*)$ appears because only CheB-P free in solution is dephosphorylated, and not the ones bound to the receptors, and this factor gives the fraction of CheB-P that is free in solution. Since the phosphorylation-dephosphorylation reactions are very fast, one may infer the probability $a_p$ from steady state conditions: $a_p = \omega_p/(k_p B_0 + k'_p Y_0)$, where $Y_0$ is the concentration of free CheY in solution, $k'_p$ is the second order rate constant for its binding to phosphorylated CheA, and $\omega_p$ is the rate of auto-phosphorylation of active CheA.

Finally, the concentration of active receptor-CheA complexes is given by

$$A^* = A_0 \sum_{m=0}^{m_{\max}} a_m X_m \qquad . \qquad [5]$$

Equations [2-5] form a complete set of equations, which predicts the kinetics of evolution of receptor activity in terms of the activation probabilities $a_m$. The activation probabilities also depend on whether the receptor is liganded or not. It has been reported that, for small ligand concentrations, this probability may be well approximated by $a_m = m/m_{\max}$ (Sourjik and Berg, 2002; Kollman et. al., 2005). Barkai and Leibler showed that perfect adaptation can be achieved if the boundary values remain fixed at two extreme values at all concentrations: i.e., $a_0 = 0$ and $a_{m_{\max}} = 1$ independent of $L$ (Barkai and Leibler, 1997). We, therefore conjecture that

$$a_m(L) \approx \frac{m}{m_{\max}} - \gamma \frac{L}{K_L} \qquad \text{for } 1 \leq m < m_{\max}, \qquad [6]$$

where $K_L$ is the dissociation constant for the ligand-receptor binding reaction and $\gamma$ is a dimensionless constant which needs to be determined.

**The reduced model**

The model described above can be considerably simplified, without losing the essential qualitative features and without significant loss in the quantitative features, if the three intermediate methylation states between 0 and $m_{\max} = 4$ are *collapsed* into one, with activation probability fixed at the average of the three values. In essence, this means that we consider three methylation states, m=0, 1 and 2 with $a_0 = 0$ and $a_2 = 1$ at all ligand concentrations, while $a_1 = 1/2 - \gamma L/K_L$ at small ligand concentrations. As will be seen shortly, this approximation considerably simplifies the mathematical equations and permits a completely analytical solution to the problem of computing the drift velocity.

Using Eq.5, we may define the active fraction of CheA

$$a^* = \frac{A^*}{A_0} = a_1(L)X_1 + X_2 \quad [7]$$

In order to calculate the response characteristics, it is convenient to assume that the system has reached equilibrium in the absence of external attractant, which is then introduced at time $t = 0$. Let us now define consequent small perturbations in the equilibrium values as follows:

$$X_m(t) = X_m^{(0)} + \delta X_m(t)$$
$$a^*(t) = a^{*(0)} + \delta a^*(t) \quad [8]$$
$$a_1(t) = a_1^{(0)} + \delta a_1(t) = \frac{1}{2} - \gamma \frac{\delta L(t)}{K_L}$$

where $\delta L(t) = \dot{L}(t)\delta t$, with $\dot{L}(t)$ being the rate of increase of the attractant concentration. From Eq.2 and 5, it can now be shown that

$$\frac{d\delta a^*}{dt} = X_1^{(0)} \frac{d\delta a_1}{dt} + \frac{1}{2}\left[(1-a^{*(0)})\delta r - a^{*(0)}\delta b - (r^{(0)} + b^{(0)})\delta a^*\right] \quad [9]$$

Note that, in agreement with the assumption of perfect adaptation, the rate of change of the active receptor fraction depends only on the time derivative of the change in the activation probability (and therefore the attractant concentration) and not on the activation probability itself.

From Eq.3 and 4, the quantities $\delta r$ and $\delta b$ themselves depend on $\delta a^*$ as follows:

$$\delta r = \frac{w_r R_0 A_0}{(K_r + A_0(1-a^{*(0)}))^2} \delta a^*$$
$$\delta b = -\frac{w_b B'^{(0)} A_0}{(K_b + A_0 a^{*(0)})^2} \delta a^* + \frac{w_b \alpha A_0}{(K_b + A_0 a^{*(0)})} \int_0^t \delta a^*(\tau) e^{-\beta(t-\tau)} d\tau \quad [10]$$

where $\alpha = k_p a_p B_0 \left[1 + A^{*(0)}/(K_b + A^{*(0)})\right]$. Eq. 9 and 10 can now be solved together using the Laplace transform technique. The result is

$$\delta\tilde{a}^*(s) = \tilde{\chi}_a(s)\delta\tilde{a}_1(s), \text{ with } \tilde{\chi}_a(s) = \frac{X_1^{(0)}s}{s + \frac{\lambda}{s+\beta} + r^{(0)}c_1 + b^{(0)}c_2}, \quad [11]$$

where

$$\beta = d_b \frac{K_b}{K_b + A^{*(0)}} \text{ and } \lambda = \frac{\omega_b \alpha A^{*(0)}}{2(K_b + A^{*(0)})}. \quad [12]$$

We have also defined two dimensionless constants

$$c_1 = \frac{1}{2}\left[1 - (1-a^{*(0)})\frac{A^{(0)}}{K_r + A^{(0)}}\right], \quad c_2 = \frac{1}{2}\left[1 - a^{*(0)}\frac{A^{*(0)}}{K_b + A^{*(0)}}\right]. \qquad [13]$$

Eq.11, when expressed in real-time is a standard linear response relation of the form $\delta a^*(t) = \int_0^t \chi_a(t-t')\delta a_1(t')dt'$, and $\chi_a(t)$ gives the linear response function of receptor activity with the change in the activation probability (and hence the change in attractant concentration) acting as the corresponding stimulus.

Explicit inversion of the Laplace transform in Eq.11 shows that the response function has the following form:

$$\chi_a(t) = X_1^{(0)}\left[\delta(t) + e^{-\theta t}\left(\beta \cosh\beta't - \frac{\theta\beta + \theta^2 - \beta'^2}{\beta'}\sinh\beta't\right)\right], \qquad [14]$$

where $\theta = (\beta + r^{(0)}c_1 + b^{(0)}c_2)/2$ and $\beta' = \sqrt{(\beta + r^{(0)}c_1 + b^{(0)}c_2)^2/4 - \lambda}$. It is often convenient to separate the singular part and write $\chi_a(t) = X_1^{(0)}\delta(t) + \chi_a'(t)$. Fig.1 shows a plot of $\chi_a'(t)$ versus time, using parameter values in Table 1. In particular, we find $\beta \approx 0.3103\,\text{s}^{-1}$, $\theta \approx 0.1625\,\text{s}^{-1}$ and $\beta' \approx 0.057\,\text{s}^{-1}$, using these values.

It may also be confirmed that $\int_0^\infty \chi_a(t)dt = 0$, in conformity with the requirement of perfect adaptation, which is a feature of the Barkai-Leibler model. This may also be seen directly from Eq.11, by putting s=0.

**From kinase activity to tumbling regulation**

As explained in the introduction, changes in the activity of the kinase CheA are directly coupled to the flagellar motors, because CheA also phosphorylates CheY (in addition to CheB), and the phosphorylated CheY binds to the base of the flagellar motors and induces a change in the direction of rotation. It has also been shown experimentally that the flagellar motor is ultra-sensitive with respect to CheY-P; the probability of rotating clockwise, the CW bias, can be well approximated by a Hill-type expression with an exponent close to 10 (Cluzel et. al., 2000). Accordingly, we assume the following form for the rate of CCW-CW switching of the motor:

$$R_- \propto Y'^H \quad \text{with } H \approx 10 \qquad [15]$$

where $Y'$ is the concentration of CheY-P in solution. CheY-P is dephosphorylated by CheZ in solution, with a rate $k_z$. We may, therefore, write the following equation for the rate of change of $Y'$:

$$\frac{d\delta Y'}{dt} = k'_p Y_0 a_p A_0 \delta a^*(t) - k_z \delta Y'(t) \qquad [16]$$

This equation has the formal solution $\delta Y'(t) = k'_p Y_0 a_p A_0 \int_0^t e^{-k_z(t-t')} \delta a^*(t') dt'$. If we further assume that $k_z$ is much larger than the rates appearing in the exponential factors in Eq.13, a quasi-steady state approximation may be used and we arrive at

$$\delta Y'(t) \approx \frac{k'_p Y_0 a_p A_0}{k_z} \delta a^*(t) \qquad [17]$$

It should be noted, however, that this equation (as well as the following Eq.17) is valid only at times $t \gg k_z^{-1}$. From Eq.14, the corresponding change in the switch rate is given by

$$\delta R_- = H R_-^{(0)} \frac{\delta Y'(t)}{Y'^{(0)}} \qquad [18]$$

where $R_-^{(0)}$ is the switch rate in steady state, in the absence of attractant.

Let us now consider the bacterium as swimming in a 2-d plane (say, x-y), with a source of nutrient along the axis x=0 and a constant, steady gradient $L_x$ parallel to the x-axis. Let $\theta_n$ be the angle made by the direction of motion of the bacterium on the positive x-axis, after the $n$'th tumbling event. Between the n'th and n+1'th tumbling events, therefore, the bacterium experiences an effective change in the attractant concentration given by

$$\dot{L}_n^{\text{eff}} = -v L_x \cos\theta_n, \qquad [19]$$

where $v$ is the velocity of smooth swimming of the bacterium between tumbles (of the order of 10μm/s). The probability of a tumble during the time interval $[\tau : \tau + \Delta\tau]$, with the last tumble (say, the n'th one) having taken place at t=0, is given by $P(\tau)\Delta\tau$, where $P(\tau) = R_-(\tau)\exp\left(-\int_0^\tau R_-(\tau')d\tau'\right)$, and the mean run interval until the next tumble takes place is $\langle\tau\rangle_n = \int_0^\infty P(\tau)\tau d\tau$. The change in the mean tumble interval, corresponding to a change in the swich-rate is given by

$$\delta\langle\tau\rangle_n = \int_0^\infty d\tau\, \tau\, \delta R_-(\tau) e^{-R_-^{(0)}\tau} - R_-^{(0)} \int_0^\infty d\tau\, \tau e^{-R_-^{(0)}\tau} \int_0^\tau \delta R_-(t)dt \qquad [20]$$

Using Eq.11, 17 and 18, it can be shown, after some algebraic manipulations, that

$$\delta\langle\tau\rangle_n = H\tilde{\chi}_a(s=R_-^{(0)})\frac{(\dot{L}_n^{\text{eff}}/K_L)}{\left[R_-^{(0)}\right]^2 a^{*(0)}} \qquad [21]$$

where $\dot{L}_n^{\text{eff}}$ is the effective time-derivative of the attractant concentration experienced by the bacterium during the run in question.

The displacement of the bacterium in the –x direction (towards the attractant source) after N tumbles is given by

$$\Delta x_N = \sum_{k=1}^{N} v \langle \tau \rangle_k \cos \theta_k \qquad [22]$$

where $\langle \tau \rangle_k = (R_-^{(0)})^{-1} + \delta \langle \tau \rangle_k$. If we now take an average of all possible directions (assuming that the direction of run after each tumble is chosen completely randomly), using Eq.19 and Eq.21, we find that

$$\langle \Delta x_N \rangle = -\frac{v^2 \sigma}{2} \frac{(L_x / K_L)}{[R_-^{(0)}]^2} N \quad , \qquad [23]$$

where $\sigma = H \tilde{\chi} [R_-^{(0)}] / a^{*(0)}$ is a dimensionless constant, and we have used the formulae $\langle \cos \theta \rangle = 0$, while $\langle \cos^2 \theta \rangle = 1/2$. The number of tumbles over a time interval $T$ is approximately $N \approx R_-^{(0)} T$. After substituting into Eq.23, and dividing by the time interval, we obtain the drift velocity:

$$v_d = -\frac{\langle \Delta x_N \rangle}{T} = \frac{v^2 \sigma}{2} \frac{(L_x / K_L)}{R_-^{(0)}} = v \frac{\sigma}{2} l_{run} \frac{L_x}{K_L} \quad , \qquad [24]$$

where we have identified the un-perturbed mean run-length $l_{run} = v / R_-^{(0)}$. Upon comparing Eq.24 with Eq.1, we may therefore conjecture (within the assumptions of our model) that

$$\kappa(L) \approx v \frac{\sigma}{2} \frac{l_{run}}{K_L} + O\left(\frac{L}{K_L}\right), \qquad [25]$$

for small attractant concentrations. Our form for the drift coefficient in Eq.25 is similar to that in an earlier calculation by de Gennes (de Gennes, 2004), in that both are proportional to the Laplace transform of a response function evaluated at the tumbling frequency.

In order to complete our analysis, we need to calculate the numerical factor $\sigma$. As mentioned, the Hill coefficient $H \approx 10$, as has been directly measured experimentally. We assume that $\gamma \sim 1$, though this needs to be confirmed by a more detailed analysis of the equilibrium between binding/activation processes. To calculate the remaining quantities, we use the parameter values listed in Table I. The active receptor-kinase fraction $a^{*(0)}$ is found by solving Eq.2 and 3 together implicitly under steady state conditions, and we find numerically that $a^{*(0)} \approx 0.23$. Finally, we compute $\tilde{\chi}_a(R_-^{(0)})$ at $R_-^{(0)} \approx 0.5 \text{ s}^{-1}$ (which corresponds to a run time of 2s), and the numerical value turns out to be ~ 0.2838. Putting together, these values give $\sigma \approx 12$. We note, interestingly, that the large value of the Hill coefficient of the flagellar motor effectively amplifies the drift velocity by an order of magnitude. In particular, we see that a 10% difference in relative concentration (as a fraction of dissociation constant) across a run-length is predicted to produce a drift velocity about half of that of the swim-speed.

**Conclusions**

Despite using several assumptions, most notably that of negligibly small background attractant concentration and small gradients, our principal prediction for the drift velocity should be testable in experiments. In particular, it is remarkable (though, in hindsight, natural) that the drift velocity as a fraction of the swimming speed during straight runs, is proportional to the fractional difference in attractant concentration over a run length (though, here, the fraction has to be calculated using the dissociation constant of the attractant-receptor binding, and not the mean concentration). The finer details of the biochemical network appear as a single dimensionless constant $\sigma$, which should be easily measurable in experiments.

At present, we are engaged in extending our calculations to non-zero background attractant concentrations. A likely modification at large concentrations is that the dissociation constant appearing in the denominator of Eq.24 might be replaced by the mean concentration. We are also interested in studying the higher order effects in terms of the attractant gradient (the present analysis was confined to the leading order). Other avenues of future exploration include the effects of noise and fluctuations and their role in determining the limit of detection, as has been discussed by Berg and Purcell (Berg and Purcell, 1977) and more recently by Bialek and Setayeshgar (Bialek and Setayeshgar, 2005).

**Acknowledgements**

MR would like to acknowledge the Harish-Chandra Research Institute, Allahabad for a summer research fellowship and hospitality during the program. MG would like to thank Mahesh Tirumkudulu and K. V. Venkatesh for sharing the details of their experiments and many fruitful discussions. We also thank the organizers of IPCMB2008 for a very stimulating inter-disciplinary meeting.

| Quantity | Symbol (this paper) | Experimental value |
|---|---|---|
| CheA concentration | $A_0$ | 5.3 µM |
| CheY concentration | $Y_0$ | 9.7 µM |
| CheR concentration | $R_0$ | 0.16 µM |
| CheB concentration | $B_0$ | 0.28 µM |
| CheR-CheA binding | $K_r$ | 0.39 µM |
| CheB-P – CheA binding | $K_b$ | 0.54 µM |
| Methylation time constant | $\omega_r$ | 0.75 s$^{-1}$ |
| Demethylation time constant | $\omega_b$ | 0.6 s$^{-1}$ |
| CheA auto-phosphorylation rate | $\omega_p$ | 23.5 s$^{-1}$ |
| CheY phosphorylation rate | $k'_p$ | 100 µM$^{-1}$ s$^{-1}$ |
| CheY-P dephosphorylation rate | $k_z$ | 30 s$^{-1}$ |
| CheB phosphorylation rate | $k_p$ | 10 µM$^{-1}$ s$^{-1}$ |
| CheB-P dephosphorylation rate | $d_b$ | 1 s$^{-1}$ |

TABLE I: A list of the experimentally measured parameters used in this paper, from Emonet and Cluzel (2008), which also gives the original references for these numbers.

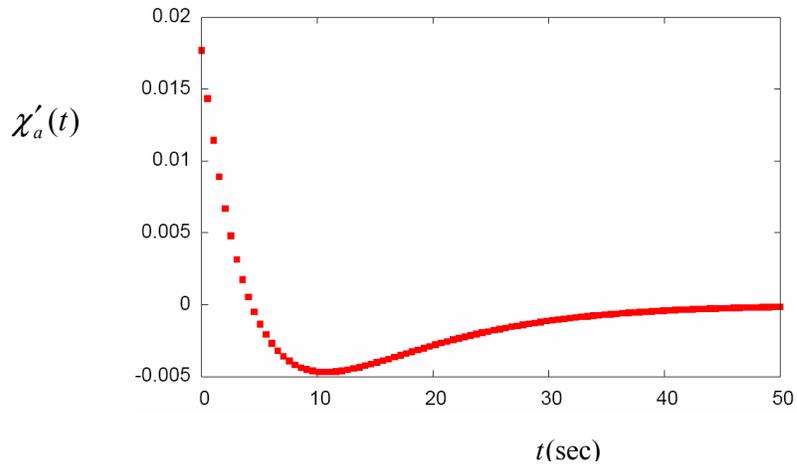

FIG 1. The figure shows the response function of the kinase activity in Eq.13, without the singular part. Note the initial sharp rise in response followed by a negative lobe, which signifies adaptation due to the CheB-dependent feedback.